\documentclass{svjour2}                    % onecolumn
\usepackage{graphicx}
\journalname{Foundation of Physics}
\begin{document}

\title{From a 1D completed scattering and double slit diffraction to
the quantum-classical problem for isolated systems
} %\subtitle{Do you have a subtitle?\\ If so, write it here}


\author{Nikolay L. Chuprikov %etc.
}


\institute{N. L. Chuprikov \at
              Tomsk State Pedagogical University, 634041, Tomsk, Russia \\
              \email{chnl@tspu.edu.ru}           %  \\
}

\date{Received: date / Accepted: date}

\maketitle

\begin{abstract}
By probability theory the probability space to underlie the set of statistical data described by the squared
modulus of a coherent superposition of microscopically distinct (sub)states (CSMDS) is non-Kolmogorovian and,
thus, such data are mutually incompatible. For us this fact means that the squared modulus of a CSMDS cannot be
unambiguously interpreted as the probability density and quantum mechanics itself, with its current approach to
CSMDSs, does not allow a correct statistical interpretation. By the example of a 1D completed scattering and
double slit diffraction we develop a new quantum-mechanical approach to CSMDSs, which requires the decomposition
of the non-Kolmogorovian probability space associated with the squared modulus of a CSMDS into the sum of
Kolmogorovian ones. We adapt to CSMDSs the presented by Khrennikov ({\it Found. of Phys., 35, No. 10, p.1655
(2005)}) concept of real contexts (complexes of physical conditions) to determine uniquely the properties of
quantum ensembles. Namely we treat the context to create a time-dependent CSMDS as a complex one consisting of
elementary (sub)contexts to create alternative subprocesses. For example, in the two-slit experiment each slit
generates its own elementary context and corresponding subprocess. We show that quantum mechanics, with a new
approach to CSMDSs, allows a correct statistical interpretation and becomes compatible with classical physics.

\end{abstract}

%\paragraph{Paragraph headings} Use paragraph headings as needed.

%%%\begin{figure}
%%%  \includegraphics{example.eps}
%%%\caption{Please write your figure caption here}
%%%\label{fig:1}       % Give a unique label
%%%\end{figure}
%%%\begin{figure*}
%%%  \includegraphics[width=0.75\textwidth]{example.eps}
%%%\caption{Please write your figure caption here}
%%%\label{fig:2}       % Give a unique label
%%%\end{figure*}
%%%\begin{table}
%%%\caption{Please write your table caption here}
%%%\label{tab:1}       % Give a unique label
%%%\begin{tabular}{lll}
%%%\hline\noalign{\smallskip}
%%%first & second & third  \\
%%%\noalign{\smallskip}\hline\noalign{\smallskip}
%%%number & number & number \\
%%%number & number & number \\
%%%\noalign{\smallskip}\hline
%%%\end{tabular}
%%%\end{table}

\newcommand{\Api}{A^{in}}
\newcommand{\Ami}{B^{in}}
\newcommand{\Apo}{A^{out}}
\newcommand{\Amo}{B^{out}}
\newcommand {\uta} {\tau_{tr}}
\newcommand {\utb} {\tau_{ref}}

\section{Introduction} \label{intro}

A 1D completed scattering and two-slit diffraction are the most simple one-particle scattering problems to arise
in quantum mechanics, and, as is commonly accepted, the contemporary quantum-mechanical models of these phenomena
give an exhaustive, maximally complete in quantum mechanics description. Moreover, these models, together with
the uncertainty and complementarity principles, as well as with Bell's classical-like analysis of the
Einstein-Podolsky-Rosen-Bohm (EPR-Bohm) experiment, are considered as essential ingredients of the modern picture
of the micro-world.

However, this picture is very strange. It allows a particle to tunnel with a superluminal velocity through an
opaque potential barrier (the Hartman paradox) and to pass simultaneously through two slits in the screen. It
discards local hidden variables (LHVs), whose existence is postulated in classical physics, and endows a particle
with mutually incompatible properties.

For a long time all ingredients of this picture have been at the heart of the long-standing debates (see, e.g.,
\cite{Bal,Hom0,Leg}) on the problem of the quantum-to-classical transition (quantum-classical problem). And all
of them eventually force researchers to discard the idealization of isolated systems and treat the
quantum-classical problem as a measurement (or macro-objectification) one, as well as to search for alternative,
"quantum" logic for interpreting micro-phenomena. Within this picture the quantum-mechanical superposition
principle conflicts with those of the macroscopic realism \cite{Leg}, and, as a consequence, quantum mechanics
looks as an ineffective theory, in which the micro-world is "{\it unspeakable}" \cite{Bell} and macro-world is
{\it undescribable}.

At the same time Bell himself believed that {\it "The quantum phenomena do not exclude a uniform description of
micro and macro worlds\dots"} \cite{Bell}. Like de Broglie, in answering the question {\it "wave {\it or}
particle?"}, he said {\it "wave {\it and} particle"} [ibid]. About this question Gerard 't~Hooft writes "Many
researchers are led to believe that the microscopic world is controlled by 'a different kind of logic' than our
classical logic. We insist that there exists only one kind of logic, even if the observed phenomena are difficult
to interpret" \cite{Hoo}.

Bohr, as is known, reduces the problem of explaining a micro-phenomenon at the micro-level to that of the
unambiguous interpretation of the corresponding {\it experimental} data. He wrote that {\it "it is decisive to
recognize that, however far the phenomena transcend the scope of classical physical explanation, the account of
all evidence must be expressed in classical terms"} \cite{Bohr}. Or, in other words, {\it "the unambiguous
interpretation of any measurement must be essentially framed in terms of the classical physical theories"} [{\it
ibid}].

However, such a "way out" raises objections. Yes, Bohr, with his brilliant intuition, truly evaluated the key
role of classical theories in studying any quantum phenomenon. But his requirement is only a part of truth,
because the unambiguous interpretation of the corresponding experimental data must be also framed in terms of
that quantum model to describe this phenomenon. Thus, the unambiguous interpretation of such data implies also
the compatibility of quantum and classical logics (in line with \cite{Hoo}).

At the same time, neither Bell's theory of LHVs and the above two models of one-particle phenomena where the
particle's state represents a CSMDS, nor the logic to underlie the uncertainty principle (the widespread
interpretation of Heisenberg's uncertainty relation) obey this requirement. Both, Bell's model of LHVs (see
\cite{Hre}) and the uncertainty principle (see \cite{Bal}), are at variance with probability theory. As regards
the current approach to CSMDSs, it opposes the quantum-mechanical superposition principle not only to the laws of
the macro-world, but also to Born's interpretation of the squared modulus of the wave function.

Indeed, the squared modulus of a CSMDS describes statistical data which do not belong to a single Kolmogorovian
probability space (see \cite{Ac1}). Hence such data are mutually incompatible (see \cite{Hr1}), what means that
the squared modulus of a CSMDS cannot be unambiguously treated as the probability density and quantum mechanics
itself, with its current approach to CSMDSs, cannot be endowed with a correct statistical interpretation.

At the same time, as is shown in \cite{Ch26,Ch27,Ch3,Ch29} by the examples of a 1D completed scattering and
two-slit diffraction, quantum mechanics allows another approach to CSMDSs, which reconciles the superposition
principle with the laws of the macro-world and corrects Born's interpretation in the case of CSMDSs. Within this
approach the corpuscular and wave properties of a particle are compatible with each other, and the standard
concepts of the group and dwell times do not lead to the Hartman paradox in studying the temporal aspects of the
tunneling phenomenon.

The aim of this paper is to analyze in detail all these questions, to join the main idea of the approach
\cite{Ch26,Ch27,Ch3,Ch29} to CSMDSs with those to underlie the generalized probabilistic approach \cite{Hr1}, and
to present on this basis a new picture of the micro-world, in which quantum mechanics allows a correct
statistical interpretation and, hence, is compatible with classical physics.

The plan of this paper is as follows. In Sects.~\ref{a1} and \ref{slit} we present, respectively, the novel
quantum-mechanical models of a 1D completed scattering and two-slit diffraction. In Sect.~\ref{a4} we discuss the
program of solving the quantum-classical problem for isolated systems, based on these two models and the
generalized probabilistic approach \cite{Hr1}.

\newcommand{\ppp}{\mbox{\hspace{5mm}}}
\newcommand{\ooo}{\mbox{\hspace{3mm}}}
\newcommand{\ooa}{\mbox{\hspace{1mm}}}
\newcommand{\ppa}{\mbox{\hspace{25mm}}}
\newcommand{\ppb}{\mbox{\hspace{35mm}}}
\newcommand{\ppc}{\mbox{\hspace{10mm}}}
\section{A 1D completed scattering as combination of two coherently evolved subprocesses -- transmission and
reflection} \label{a1}
\subsection{Backgrounds} \label{a10}

The main motive to revise the contemporary model of a 1D scattering (CMS) is that the group and dwell times
aimed, within the CMS, to describe the duration of the tunneling phenomena saturate with increasing the width of
opaque potential barriers (the usual Hartman paradox) and even with increasing the space between such barriers in
many-barriers structures (the generalized Hartman paradox) (see, e.g., \cite{Har,Ol3,Ol2,Win,Lun,Nim}). What is
important is that this paradox does not disappear in the relativistic approach (see \cite{Lun,Krek,Zhi}).

Unlike \cite{Win,Nim} to consider the above saturation as a real physical {\it effect} to demand a proper
interpretation, we do it as a {\it real paradox} to reveal the existence of a serious defect in the CMS. The main
reason is that introducing the tunneling time needs the knowledge of the whole time evolution of transmitted
particles within the barrier region. However, the CMS where {\it "\ldots an incoming peak or centroid does not,
in any obvious physically causative sense, turn into an outgoing peak or centroid\ldots"} \cite{La2,La1} does not
obey this requirement. The Hartman paradox results from ignoring this requirement within the CMS, in introducing
the tunneling time.

To present the main idea of the approach \cite{Ch26,Ch27,Ch3}, we consider here the stationary 1D scattering
problem. Its setting is as follows. A particle with energy $E$ impinges from the left the symmetrical potential
barrier $V(x)$, $V(x-x_c)=V(x_c-x)$, confined to the finite spatial interval $[a,b]$; $d=b-a$ is the barrier
width; $x_c$ is the midpoint of the barrier region. For this setting the particle's state $\Psi_{full}(x;k)$
($k=\sqrt{2mE}/\hbar$) can be written as follows
\begin{eqnarray} \label{511}
\Psi_{full}(x;k)=\left\{ \begin{array}{c} e^{ikx}+b_{out}(k)e^{ik(2a-x)}\ppa x\le a\\
a_{full} F(x-x_c;k)+b_{full} G(x-x_c;k) \ppp a\le x\le b\\
a_{out}(k)e^{ik(x-d)}\ppb x>b
\end{array}\right.
\end{eqnarray}
$F(x-x_c;k)$ and $G(x-x_c;k)$ are such real solutions to the Schr\"odinger equation that
$F(x_c-x;k)=-F(x-x_c;k)$, $G(x_c-x;k)=G(x-x_c;k)$; their (constant) Wronskian $\frac{dF}{dx}G-\frac{dG}{dx}F$
will be further denoted via $\kappa$;
\begin{eqnarray} \label{51300}
a_{out}=\frac{1}{2}\left(\frac{Q}{Q^*}-\frac{P}{P^*}\right);\ooo
b_{out}=-\frac{1}{2}\left(\frac{Q}{Q^*}+\frac{P}{P^*}\right);
\end{eqnarray}
\begin{eqnarray*}
a_{full}= -\frac{1}{\kappa}P^*a_{out}e^{ika};\ooo b_{full}=\frac{1}{\kappa}Q^*a_{out}e^{ika};\nonumber
\end{eqnarray*}
\begin{eqnarray*}
Q=\left[\frac{dF(x-x_c)}{dx}+i k F(x-x_c)\right]_{x=b};\ooa P=\left[\frac{dG(x-x_c)}{dx}+i k
G(x-x_c)\right]_{x=b}
\end{eqnarray*}

For example, if $V(x)$ is the rectangular barrier of height $V_0$ ($E<V_0$),
\begin{eqnarray*}
F=\sinh(\kappa x),\ooo G=\cosh(\kappa x),\ooo \kappa=\sqrt{2m(V_0-E)}/\hbar.
\end{eqnarray*}

\subsection{Wave functions for transmission and reflection} \label{a12}

By the approach \cite{Ch26,Ch27,Ch3} a 1D completed scattering is a complex process to consist of two coherently
evolved subprocesses -- transmission and reflection. This idea implies that the searched-for stationary wave
functions $\psi_{tr}(x;k)$ and $\psi_{ref}(x;k)$ to describe, respectively, transmission and reflection obey the
following two requirements: (\i) $\psi_{tr}(x;k)+\psi_{ref}(x;k)=\Psi_{full}(x;k)$; (\i\i) to provide for a
causal relationship between the incoming and outgoing waves, each subprocess's wave function as well as the
corresponding probability current density must be continuous at the joining point.

In the second requirement it is taken into account that there is no solution to the stationary Schr\"odinger
equation for a semitransparent potential barrier, which would be everywhere continuous together with its first
derivative and, in addition, have one incoming and one outgoing waves. As was shown in \cite{Ch26,Ch27,Ch3}, for
symmetrical barriers, the midpoint of the barrier region is a particular point for the subprocesses. It divides
the $OX$-axis into two parts, where the outgoing waves of the functions $\psi_{ref}(x;k)$ and $\psi_{tr}(x;k)$
are joined with their incoming counterparts. As a result, the wave functions for reflection and transmission read
as follows. For $x\leq a$
\begin{eqnarray} \label{700}
\psi_{ref}(x;k)=\Api_{ref}e^{ikx}+b_{out}e^{ik(2a-x)}, \ppp \psi_{tr}(x;k)=\Api_{tr}e^{ikx};
\end{eqnarray}
for $a\le x\le x_c$
\begin{eqnarray} \label{701}
\psi_{ref}(x;k)=\kappa^{-1}\left(PA^{in}_{ref}+
P^*b_{out}\right)e^{ika}F(x-x_c;k)\nonumber\\
\psi_{tr}(x;k)=\kappa^{-1}PA^{in}_{tr}e^{ika}F(x-x_c;k)+ b_{full}G(x-x_c;k);
\end{eqnarray}
for $x\ge x_c$
\begin{eqnarray} \label{2}
\psi_{ref}(x;k)\equiv 0,\ppp \psi_{tr}(x;k)\equiv \Psi_{full}(x;k);
\end{eqnarray}
\begin{eqnarray} \label{3}
\Api_{ref}=b_{out}^*\left(b_{out}+a_{out}\right);\ooo \Api_{tr}= a_{out}\left(a^*_{out}-b^*_{out}\right).
\end{eqnarray}
As it follows from Expr. (\ref{700})-(\ref{2}), the main feature of $\psi_{tr}(x;k)$ and $\psi_{ref}(x;k)$ is
that either function, containing one incoming and one outgoing wave, is continuous at the joining point $x_c$,
together with the corresponding probability current density.

So, by this approach, in the case of reflection particles never cross the point $x_c$. This result agrees with
the well known fact that, for a classical particle to impinge from the left a smooth symmetrical potential
barrier, the midpoint of the barrier region is the extreme right turning point, irrespective of the particle's
mass and the barrier's form and size. That is, by the classical logic, in the regions $x<x_c$ and $x>x_c$ a
particle moves under different physical contexts (see also Sect.~\ref{a4}). This implies that in these regions
either subprocess is described by different (properly joined at the point $x_c$) solutions of the Schr\"odinger
equation, resulting in the unusual (piecewise continuous) character of the wave functions to describe the
subprocesses.

Now, before presenting characteristic times to describe the transmission and reflection of a particle with a
given energy $E$, we will dwell shortly on some peculiarities of the subprocesses in the non-stationary case. As
is seen from (\ref{3}), not only $\Api_{tr}+\Api_{ref}=1$, but also $|\Api_{tr}|^2+|\Api_{ref}|^2=1$. From this
it follows that the narrow in $k$-space wave packets $\psi_{tr}(x,t)$ and $\psi_{ref}(x,t)$ built out,
respectively, of $\psi_{tr}(x;k)$ and $\psi_{ref}(x;k)$ with different values of $k$, obey the following
relations
\[\Re\langle\psi_{tr}(x,t)|\psi_{ref}(x,t)\rangle=0.\] Thus, despite the existence of interference between
$\psi_{tr}$ and $\psi_{ref}$, for any $t$
\begin{eqnarray*}
\langle\Psi_{full}(x,t)|\Psi_{full}(x,t)\rangle
=\textbf{T}+\textbf{R}=1;\\
\textbf{T}=\langle\psi_{tr}(x,t)|\psi_{tr}(x,t)\rangle;\ooo
\textbf{R}=\langle\psi_{ref}(x,t)|\psi_{ref}(x,t)\rangle;
\end{eqnarray*}
$\textbf{T}$ and $\textbf{R}$ are (constant) transmission and reflection coefficients, respectively. Moreover, in
this limiting case, Ehrenfest's theorem for the expectation values $<\hat{x}>$ and $<\hat{p}>$ of the particle's
position $\hat{x}$ and momentum $\hat{p}$ holds:
\begin{eqnarray} \label{4}
\frac{d<\hat{x}>_{tr}}{dt}=\frac{1}{m}\left<p\right>_{tr} \ppp
\frac{d<\hat{x}>_{ref}}{dt}=\frac{1}{m}\left<p\right>_{ref}.
\end{eqnarray}

Note that $\textbf{R}$ remains always constant in the general case too. However $\textbf{T}$ is now constant only
in the initial and final stages, i.e., long before and long after the scattering event. In the very stage of
scattering, $d\textbf{T}/dt=I_{tr}(x_c+0,t)-I_{tr}(x_c-0,t)\neq 0$; here $I_{tr}$ is the probability current
density to correspond to $\psi_{tr}(x,t)$. (As the continuity equation for the wave function is nonlinear, the
continuity of $I_{tr}$ for each wave of the wave packet does not guarantee that the whole packet obeys this
equation.)

Thus, in the general case, for the instants of time which correspond to the very stage of scattering, the quantum
mechanical formalism does not allow one to entirely exclude the interference terms from $\psi_{tr}(x,t)$, in
partitioning the whole scattering process into alternative subprocesses. It should be stressed however that, in
our numerical calculations for wave packets whose initial width was comparable with the barrier width, the
relative deviation of $\textbf{T}$ from the constant $1-\textbf{R}$ did not exceed several percentages. (Note
that in the opposite limiting case, i.e., for wave packets narrow in the $x$-space at the initial instant of
time, the role of the joining point $x_c$ diminishes because of a quick widening of such packets. In this limit
$\textbf{T}$ tends to be constant.)

\subsection{Dwell times for transmission and reflection} \label{a13}

Now, when quantum dynamics of both the subensembles of particles - transmitted and reflected - has been revealed
in all stages of a 1D completed scattering, we can present the scattering times to characterize these
subensembles. The validity of Exps. (\ref{4}) as well as the continuity, at the point $x_c$, of the wave
functions $\psi_{tr}(x;k)$ and $\psi_{ref}(x;k)$ and the corresponding probability current densities are of
importance for a correct introduction of the group and dwell times for the subprocesses of a 1D completed
scattering.

For the details of introducing the group transmission and reflection times to describe wave packets see
\cite{Ch27,Ch3}. Here we present only the dwell times for these subprocesses. The concept of the group time is
considered in our approach as auxiliary one. For example, it becomes useless in the stationary case. Therefore
the main role in the time-keeping of a scattering particle in the barrier region is played here by the dwell time
which describes these subprocesses both in the stationary and non-stationary cases.

The transmission ($\tau_{tr}^{dwell}$) and reflection ($\tau_{ref}^{dwell}$) dwell times read as
\begin{eqnarray*}
\tau_{tr}^{dwell}(k)=\frac{1}{I_{tr}}\int_a^b|\psi_{tr}(x;k)|^2 dx,\ooo \tau_{ref}^{dwell}(k)=\frac{1}{I_{ref}}
\int_a^{x_c}|\psi_{ref}(x,k)|^2 dx;
\end{eqnarray*}
$I_{tr}=I_{full}=T(k)\hbar k/m$, $I_{ref}=R(k) \hbar k/m$; $T=|a_{out}|^2$, $R=|b_{out}|^2$.

As is shown in \cite{Ch27}, for the rectangular barrier $V_0$, for $E<V_0$
\begin{eqnarray} \label{4007}
\tau_{tr}^{dwell}(k)=\frac{m}{2\hbar k\kappa^3}\left[\left(\kappa^2-k^2\right)\kappa d +\kappa_0^2 \sinh(\kappa
d)\right];
\end{eqnarray}
\begin{eqnarray*}
\tau_{ref}^{dwell}(k)=\frac{m k}{\hbar \kappa}\cdot\frac{\sinh(\kappa d)-\kappa d}{\kappa^2+\kappa^2_0
\sinh^2(\kappa d/2)};
\end{eqnarray*}
$\kappa_0=\sqrt{2mV_0}/\hbar$.

In this case, within the CMS the dwell time $\tau_{dwell}$ (see \cite{But}) reads as
\begin{equation} \label{4009}
\tau_{dwell}(k)= \frac{m k}{\hbar \kappa}\cdot \frac{2\kappa d (\kappa^2-k^2)+\kappa_0^2 \sinh(2\kappa
d)}{4k^2\kappa^2+ \kappa_0^4\sinh^2(\kappa d)}.
\end{equation}

As is seen from (\ref{4007}), $\tau_{tr}^{dwell}$ increases exponentially rather than saturates in the limit
$d\to\infty$. Thus, in our approach, the dwell transmission time, contrary to the dwell time $\tau_{dwell}$
(\ref{4009}) introduced in the CMS, does not lead to the Hartman effect. By our approach the opaque barrier
strongly retards the motion of a particle with a given energy $E$, when it enters the barrier region.

Moreover, as is shown in \cite{Ch29} for a particle tunnelling through the system of two identical rectangular
barriers of width $d$ and height $V_0$, with the distance $l$ between them, our approach does not lead to the
generalized Hartman effect which was found in \cite{Ol3}, within the CMS. In presenting the behavior of
$\tau_{tr}^{dwell}$ and $\tau_{ref}^{dwell}$ in this limit, we have to take into account that these quantities
possess the property of additivity. They can be written as follows, \[\tau_{tr}^{dwell}=\tau^{(1)}_{tr}
+\tau^{gap}_{tr} +\tau^{(2)}_{tr},\ooo \tau_{ref}^{dwell}=\tau^{(1)}_{ref} +\tau^{gap}_{ref};\] $\tau^{(1)}_{tr}$
and $\tau^{(1)}_{ref}$ describe the first barrier; $\tau^{gap}_{tr}$ and $\tau^{gap}_{ref}$ do the gap between
the barriers, and $\tau^{(2)}_{tr}$ describes the second barrier (remind that reflected particles do not cross
the midpoint of this symmetric structure). As is shown in \cite{Ch29}, in the opaque-barrier limit, when
$V_0\to\infty$ (or $\kappa_0\to\infty$) and $d$ is fixed,
\begin{eqnarray*}
\tau^{(1)}_{tr}=\tau^{(2)}_{tr}\approx\frac{m}{4\hbar k\kappa_0}\exp(2\kappa_0d),\ooo \tau^{gap}_{tr}\approx
\frac{m\kappa_0^2}{8\hbar k^4}\left[kl-\sin(kl)\right]\exp(2\kappa_0d);\\
\tau^{dwell}_{ref}\approx\tau^{(1)}_{ref}\approx \tau_{dwell}\approx \frac{2mk}{\hbar \kappa_0^3}.
\end{eqnarray*}
As is seen, $\tau_{tr}^{dwell}$ increases exponentially in this limit, and what is also important is that it
depends on the distance $l$ between the barriers. Moreover, in the course of passing a particle through the
structure, it spends the most part of time just in the space between the opaque barriers. Contrary, the dwell
time $\tau_{dwell}$ does not depend, in this limit, on the distance between the barriers and diminishes when
$\kappa_0\to\infty$.

Analogous situation arises in another opaque-barrier limit, when $d\to\infty$ and $V_0$ is fixed. Now the
explicit expressions for the dwell times becomes somewhat complicated (see \cite{Ch29}), giving no qualitatively
new information, and we omit this case.

\section{On the compatibility of the wave and corpuscular properties of a particle in the two-slit experiment} \label{slit}

So, the main shortcoming of the conventional description of a 1D completed scattering is that within its
framework the state of a particle taking part in the process to imply for a particle two alternative
possibilities (either to be transmitted through the barrier or to be reflected by it) is a CSMDS only in the
final stage of this process. And just to show that quantum mechanics really allows the presentation of this state
as a CSMDS for all stages of this process was the main purpose of Sect.~\ref{a1}. Treating this quantum process
as a complex one to consist of two alternative subprocesses is essential for the unambiguous interpretation
(i.e., without discarding the classical logic) of all experimental data associated with a 1D completed
scattering.

At first glance, in the case of a two-slit diffraction to imply for a particle two alternative possibilities,
too, we meet a simpler situation. Within its conventional description the particle's state is a CSMDS in all
stages of this process. Moreover, in the limiting case, when slits in the screen are infinitely removed from each
other (i.e., when there is no interference between the "one-slit" states to enter the CSMDS), this description
creates no problems for the unambiguous interpretation of experimental data associated with this quantum
phenomenon.

However, for a finite distance between the slits, the norm of the CSMDS cannot be presented, because of
interference, as the sum of the norms of the one-slit states. Hence, under this condition, these states cannot be
associated with alternative subprocesses of a two-slit diffraction. Within the conventional description of this
process, this fact is usually interpreted as the incompatibility of the wave and corpuscular properties of a
particle.

In this connection, our next aim is to show, by the example of a strictly symmetrical setting of the two-slit
experiment, that the above CSMDS with the interfering one-slit states can be really transformed into that with
states to describe {\it alternative} subprocesses. The classical logic suggests us that the alternative
subprocesses, being independent for the slits infinitely removed from each other, become mutually dependent when
the distance between the slits is finite (in Sect.~\ref{a4} we associate this property with the interaction of
the elementary contexts created by each slit). Thus the wave functions to describe alternative subprocesses must
depend, unlike the one-slit states, on the distance between the slits.

In the considered setting of the two-slit experiment we assume that the $YZ$-plane coincides with the plane of
the first screen to have two parallel identical slits centered on the planes $y=-a$ (first slit) and $y=a$
(second slit), and the wave function $\Psi_{two}(x,y,z;E)$ to describe a particle with energy $E$ when both the
slits are opened has the form
\begin{equation} \label{71}
\Psi_{two}(x,y,z;E)\equiv\Phi(x,y,z;E);
\end{equation}
\[\Phi(x,y,z;E)=
\Psi_{one}(x,y-a,z;E)+\Psi_{one}(x,y+a,z;E);\] $\Psi_{one}(x,y,z;E)$ is the one-slit wave function:
$\Psi_{one}(0,y,z;E)=0$ for $|y|>d/2$; $\Psi_{one}(x,-y,z;E)=\Psi_{one}(x,y,z;E)$; $d$ is the slits' width;
$a>d/2$. It is also assumed that a particle impinges the first screen from the left, and the second screen -- the
particle' s detector -- coincides with the plane $x=L$ ($L>0$).

It is evident that in this setting
\begin{equation} \label{81}
\Psi_{two}(x,-y,z;E)=\Psi_{two}(x,y,z;E).
\end{equation}
Thus, the $y$-th components of the probability current densities associated with the first and second slits
balance each other on the plane $y=0$, and the $y$-th component of the probability current density associated
with their superposition is zero on this plane (see also \cite{Hom} where this experiment is analyzed within the
Bohmian approach). If one inserted along the symmetry plane (i.e., with keeping the condition (\ref{81})) an
infinitesimally thin two-side "mirror"\ (we use here this word, bearing in mind the analogue between this quantum
experiment and its counterpart in classical electrodynamics) to elastically scatter particles, the wave function
(\ref{71}) and hence the interference pattern to appear in this experiment on the second screen would remain the
same.

The coincidence of the particle's states in the original (without the mirror) and modified (with the mirror)
two-slit experiments leads us to the following conclusions: (\i) the ensemble of particles, freely moving between
the first and second screens in the original experiment, is {\it equivalent} to the ensemble of particles in the
modified experiment where a particle {\it a priori} cannot cross the plane $y=0$ occupied by the mirror; (\i\i)
in the original experiment a particle always passes only through one of two open slits. Figuratively speaking,
the procedure of inserting the mirror can be considered as a non-demolishing "which-way"\ measurement.

From (\i) it follows that the wave function (\ref{71}) can be rewritten as
\begin{equation} \label{73}
\Psi_{two}(x,y,z;E)=\psi_{two}^{(1)}(x,y,z;E)+\psi_{two}^{(2)}(x,y,z;E);
\end{equation}
where $\psi_{two}^{(1)}$ and $\psi_{two}^{(2)}$ are piecewise continuous wave functions --
\begin{eqnarray*}  \psi_{two}^{(1)}(x,y,z;E)=\left\{ \begin{array}{c}
0, \ppa y>0\\
\Phi(x,y,z;E), \ppc y<0
\end{array}\right. \\
\psi_{two}^{(2)}(x,y,z;E)=\left\{ \begin{array}{c} \Phi(x,y,z;E),\ppc y>0\\
0, \ppa y<0
\end{array}\right.
\end{eqnarray*}
$\psi_{two}^{(1)}(x,0,z;E)=\psi_{two}^{(2)}(x,0,z;E)=\Phi(x,0,z;E)/2$. We have to stress that these expressions
are valid both for $x>0$ and for $x<0$.

Though Exps. (\ref{71}) and (\ref{73}) give the same wave function, the physical meaning of
$\Psi_{one}(x,y-a,z;E)$ and $\Psi_{one}(x,y+a,z;E)$ to enter (\ref{71}) differs cardinally from that of
$\psi_{two}^{(1)}(x,y,z;E)$ and $\psi_{two}^{(2)}(x,y,z;E)$ to enter (\ref{73}). The last pair describes the {\it
alternative} subprocesses of the two-slit diffraction, as
\begin{equation} \label{74}
\|\Psi_{two}\|^2=\|\psi_{two}^{(1)}\|^2+\|\psi_{two}^{(2)}\|^2;
\end{equation}
here $\|\Psi\|^2=\int_{-\infty}^\infty |\Psi(L,y,z;E)|^2dy$ for any value of $z$; since the structure considered
is uniform in the $z$-direction, this norm does not depend on $z$. At the same time, the first pair cannot be
associated with alternative subprocesses, as $\|\Psi_{two}\|^2\neq\|\Psi_{one}(x,y-a,z;E)\|^2+
\|\Psi_{one}(x,y+a,z;E)\|^2$.

CSMDS (\ref{73}) obeys the "either-or" rule to govern alternative random events in classical probability theory.
In line with (\ref{74}), the whole ensemble of particles described by (\ref{73}) consists of two subensembles:
one of them, described by $\psi_{two}^{(1)}$, consists of particles to pass through the first slit, provided that
the second one is opened; another subensemble, described by $\psi_{two}^{(2)}$, consists of particles to pass
through the second slit, provided that the first one is opened. Each subensemble creates an {\it unremovable}
context for its counterpart (in more detail, the notion 'context' will be analyzed in Section \ref{a45}).

So, interference plays a twofold role in the two-slit experiment. On the one hand, namely interference makes it
{\it impossible} to associate the one-slit wave functions $\Psi_{one}(x,y-a,z;E)$ and $\Psi_{one}(x,y+a,z;E)$
with alternative subprocesses of a two-slit diffraction. On the other hand, namely interference makes it {\it
possible} to decompose a two-slit diffraction into two alternative subprocesses described by the "one-subprocess"
states $\psi_{two}^{(1)}$ and $\psi_{two}^{(2)}$ to be inseparable from each other.

\section{About the origin of the quantum-classical problem and its resolution on the basis of
a new approach to CSMDSs} \label{a4}

So, from the mathematical viewpoint the main innovation in our approach to a 1D completed scattering and two-slit
diffraction is the representation of the wave function of either process in the form of a CSMDS whose norm is
equal to the sum of norms of its substates. The only exclusion appears in the case of a 1D completed scattering,
for wide (in the momentum space) wave packets, in the very stage of the scattering event. From the physical
viewpoint its innovation is that it treats these two quantum one-particle phenomena as {\it complex} processes,
i.e., as those consisting of alternative subprocesses (again, except for the above case).

Both the quantum-mechanical models are consistent with the classical logic. They imply that a micro-particle,
like a macro-particle, can pass only through one of two open slits in the screen, as well as can {\it either} be
transmitted {\it or} reflected by a potential barrier. Thus, by this approach the abbreviation "CSMDS" is equally
applicable both to {\it micro-} and {\it macro-}particles.

Of course, such conformity of quantum-mechanical models with classical physics contradicts Bell's analysis of
LHVs as well as the uncertainty and complementarity principles. However, neither Bell's analysis (see review
\cite{Hre}) nor the uncertainty principle (see \cite{Bal}) and complementarity principle (see \cite{Hr1}) reflect
properly the inherent properties of micro-particles. Our next step is to consider these questions in detail and
to show that quantum mechanics, based on the ideas of contextuality and supplementarity developed within the
"contextual statistical realist model" \cite{Hr1} as well as on our approach to CSMDSs, allows a correct
statistical interpretation and gets free of the quantum-classical problem for isolated systems.

\subsection{On Schr\"odinger's and Bell's visions of the quantum-classical problem} \label{a41}

Originally, mainly due to Schr\"odinger's cat paradox, by the quantum-classical problem was meant the conflict to
arise within the conventional description of CSMDSs, when their properties were extended onto macro-systems,
between the quantum-mechanical superposition principle and the "either-or"\ rule to govern mutually exclusive
random events in classical probability theory. Later this problem was meant as the conflict to arise within
Bell's theory of EPR-Bohm's experiment between the nature of the micro-world and the classical postulate on the
absolute existence of LHVs. So that the famous Schr\"odinger's cat paradox and Bell's model of LHVs are the most
important milestones in becoming the modern vision of this problem. Both demonstrate the existence of a deep
contradiction between the quantum and classical laws.

Figuratively speaking, on the road between the micro- and macro-scales Schr\"odinger and Bell go in the opposite
directions. Schr\"odinger demonstrates the emergence of this conflict when one attempts to extend the quantum
laws onto the macro-scales -- he calls in question the validity of the superposition principle at the
macro-scales. Bell shows its emergence when one attempts to extend the classical laws onto the atomic scales --
he calls in question the universal validity of such fundamental notion of classical physics as "causal external
world", i.e., the world to exist independently of an observer and to obey the principles of relativity theory.

Note that, since quantum mechanics was invented by its founders as a {\it universal} theory, the
quantum-classical problem can be also treated as the problem of the (in)completeness of quantum mechanics. From
this viewpoint, Schr\"odinger calls in question the {\it completeness} of quantum mechanics at the {\it
macro}-scales, Bell calls in question its {\it incompleteness} at the {\it atomic} scales.

The most of scientists dealing with the quantum-classical problem treats it as the measurement (or
macro-objectification) problem whose resolution is impossible within the idealization of isolated systems. This
viewpoint is based on Bohr's principle of complementarity and Heisenberg's uncertainty principle, as well as on
the observed violation of Bell's inequality, which is considered as the falsification of Bell's assumption on the
existence of LHVs for the EPR-pairs.

Within this picture, solving the quantum-classical problem to arise for CSMDSs is impossible without suggesting
that, apart from the potential to enter the Hamiltonian of a quantum system, there is {\it "everything else"}
(e.g., the mechanisms of decoherence or localization) to affect the system's dynamics, reducing its state - a
CSMDS - to a definite substate of the CSMDS (the review of programs based on this idea is done in \cite{Sch} (see
also \cite{Joo,Ghi})).

Of course, extending quantum theory onto open systems is of importance, as there are many interesting physical
problems when the influence of an environment on systems is essential. However there is a reason to believe that
the quantum-classical problem is not a measurement one. Apart from the internal flaws of these programs (see,
e.g., \cite{Bal,Hom0,Bal1}), the very idea of the nonexistence of LHVs, which underlies these programs, has a
questionable ground.

\subsection{On the critique of Bell's theory of LHVs} \label{a42}

Let us dwell on the approaches to criticize Bell's proof of the nonexistence of LHVs for the EPR-pairs in the
thought EPR and EPR-Bohm experiments. There are three aspects of this proof, which are subjected to their
critique.

The first one is the correctness of the mathematical setting of the space frame where LHVs associated with these
experiments should "exist". For example, as is pointed out in \cite{Vo1,Vo2}, in order to judge on the
(non)existence of (non)locality in these experiments, the continuous LHVs must be, at least, embedded into the
space-time structure, being correct from the viewpoint of relativity theory. The correct embedding is shown to be
sufficient for explaining the original EPR experiment.

As regards the (two-valued) spin LHV used in studying the EPR-Bohm experiment, Bell embedded it into the
topological space $S^0$. However, as is shown in \cite{Joy}, {\it "EPR[-Bohm] elements of reality are points of a
2-sphere [$S^2$], not 0-sphere [$S^0$] as Bell assigned"}. Figuratively speaking, the two-valued spin components,
as elements of reality, cannot belong to $S^0$ where "there are no room" for distinguishing the experimental
contexts associated with the {\it differently oriented} polarization beam splitters used for detection of
electrons.

Other two aspects of Bell's proof are revealed within the approaches to criticize the derivation of Bell's
inequality from the viewpoint of quantum mechanics and probability theory (see the pioneer works by Fine
\cite{Fin}, Accardi \cite{Acc}, Pitowsky \cite{Pit} and Rastall \cite{Ras}, as well as the recent review
\cite{Hre} and papers \cite{Hes,Nie,An1}).

As was shown by Fine \cite{Fin}, apart from Bell's {\it explicit} assumption on the existence of LHVs, the
derivation of Bell's inequality is based also on the {\it implicit} assumption that there is a compatible joint
distribution to describe experimental data obtained in the EPR-Bohm experiments with the different orientations
of polarization beam splitters. From the viewpoint of quantum mechanics this assumption is improper {\it a
priori}, because such data are obtained in fact for noncommuting observables and, thus, there is no compatible
joint quantum distribution to describe the whole set of such data.

Lastly, the above implicit assumption made within the classical-like analysis is at variance with classical
probability theory (see the review \cite{Hre}) where Bell's type inequalities have been known yet before Bell,
and their violation means simply that they contain {\it incompatible} probabilities -- they describe statistical
data which cannot be associated with a single Kolmogorovian probability space. As was shown in \cite{Ac2}, {\it
"\ldots Bell's "vital [explicit] assumption" not only is not "vital" but in fact has nothing to do with Bell's
inequality\ldots this inequality cannot discriminate between local and non local hidden variable theories"}. By
\cite{An1}, {\it "Strictly speaking, there does not exist [Bell's] inequality such that all the three means
involved in it would be spin correlations. It is therefore meaningless to speak of verification of [this]
inequality\ldots Local hidden variables and probabilities, consistently introduced for the experiments to test
Bell's inequality, obey the inequality to differ from Bell's one"} (see also \cite{Joy,Hes,Hr6}).

Thus, the main lesson to follow from the above critique is that the experimental violation of Bell's inequality
does not falsify the existence of LHVs and does not say about a nonlocal character of quantum mechanics. Rather
it falsifies the existence of "Bell's LHVs"\ to be at variance with probability theory. As was shown (see, e.g.,
\cite{Ac1}), this concerns all Bell's-type "no-go" theorems, with inequalities or without theirs. And what is
also important is that, in describing the EPR-Bohm experiment, quantum mechanics and classical probability theory
respect each other, because introducing a {\it common} LHV for different experimental setups contradicts both
these theories.

Note, the story of LHVs, apart from the critique of their Bell model, contains also the positive part. During the
last three decades a careful analysis of possible ways of a causal realistic description of the micro-world has
been carried out within the generalized probabilistic approaches (see \cite{Hr5,Ac5}). Besides, the variety of
prequantum approaches (see, e.g., \cite{Hr2,HesD,Hof,Sla} and references therein) aimed to reveal the internal
structure of micro-particles and to reproduce their quantum dynamics has been developed.

\subsection{On the supplementarity principle for the particle's position and momentum} \label{a43}

Of importance for understanding the nature of quantum probabilities is the generalized probabilistic model
\cite{Hr1} "{\it \ldots based on two cornerstones: (a) contextuality of probabilities; (b) the use of two fixed
(incompatible) physical observables in order to represent the classical contextual probabilistic model in the
complex Hilbert space.}"

By the notion of a 'context' is meant here {\it a complex of physical conditions} (which can be created, in the
general case, without the participation of an observer) to "prepare" uniquely a quantum statistical ensemble and
to govern uniquely its time evolution (see \cite{Hr1} and references therein). As is said in \cite{Hr1},
"[identical] physical systems interact with a context\ldots and in this process a statistical ensemble \ldots is
formed".

By two incompatible physical ('reference') observables are meant the particle's position $x$ and momentum $p$. As
is said in \cite{Hr1}, {\it "Values associated to the \ldots $x$ and $p$ \ldots are considered as objective
properties of physical systems. The reference observables are therefore not contextual in the sense of Bohr's
measurement contextuality. However, the probabilistic description is possible only with respect to a fixed
context \ldots For a given context, [position and momentum] produce supplementary statistical information; in the
sense that the contextual probability distribution of [position] could not be reconstructed on the basis of the
probability distribution of [momentum, and vice versa]"}.

This is the essence of the so called "supplementarity" principle introduced in \cite{Hr1} instead of Bohr's
complementarity principle. As is emphasized in \cite{Hr2}, {\it "On the one hand, we deny Bohr's philosophic
principle on completeness of QM. On the other hand, we deny naive Einsteinian realism-assigning results of all
possible observations to a hidden variable\ldots Opposite to Einstein, [our theory] does not assign results of
quantum observations directly to hidden variables. Bohr's idea that the whole experimental arrangement should be
taken into account is basic for [it]\ldots"}

For a proper understanding of the difference between the supplementarity and complementarity principles it is
important to distinguish between the above mentioned context to uniquely "prepare" a quantum process and the
context created by the measuring device. The former, which includes initial conditions and external potential
fields to govern the ensemble's dynamics, enters into the quantum-mechanical description of the process. While
the latter is beyond the scope of this description. The former will be referred to as the "preparation context"
($\cal{P}$-context), the latter is the "measuring context" ($\cal{M}$-context).

If some probabilistic model introduces probabilities depending on the characteristic of the {\it measuring}
device, then the corresponding $\cal{M}$-context should be considered  as a part of the $\cal{P}$-context. Just
such situation appears in Bell's model to involve probabilities depending on the orientations of polarization
beam splitters of {\it detecting} devices. That is, the LHV introduced by Bell corresponds to the "mixture" of
incompatible $\cal{P}$-contexts.

Note that ideas like \cite{Hr1,Hr5} underlie also the approach \cite{Ac1,Acc,Ac2,Ac5}. Both generalized
probabilistic approaches present a single particle as an observer-independent entity, and both assume that the
conjugate variables, $x$ and $p$, are not {\it ontic} ones. For a single particle their values are assumed to
emerge in the course of the measuring process, i.e., in the course of the interaction of a particle with a
measuring device. Thus, the values of $x$ and $p$, measured for a single particle in a single experiment, depend
on the $\cal{M}$-context.

This assumption agrees with the prequantum models \cite{Hes,Hr2,HesD,Hof,Sla} by which these two variables to
describe exhaustively the state of a classical point-like particle cannot properly characterize the "hidden"
state of a quantum particle which seems to be not a point-like object (and not a wave packet) -- instead of the
"hidden one-particle variables" $x$ and $p$ it is better to speak of the "hidden particle's structure".

However our aim is to analyze the particle's dynamics within the scope of quantum mechanics. Therefore we have to
deal with the ensembles of particles, rather than with a single particle. The main difference between these two
cases is that, with respect to the particle's ensemble (created by some $\cal{P}$-context), it is relevant to
treat $x$ and $p$ as supplementary one-particle variables to describe ontic (predetermined by the
$\cal{P}$-context), simultaneously existing, properties of this ensemble (see also \cite{Hr1}). As is stressed in
\cite{Hr1}, "Pairs of supplementary observables produce \ldots rough (statistical) images of the underlying
reality\ldots" (and these images are unique for given $\cal{P}$-contexts.)

To exemplify all the above, let us consider the time-dependent state-vector $|\Psi\rangle$ to describe some
one-particle process (now it is important to assume that $|\Psi\rangle$ is not a CSMDS). By \cite{Hr1},
$|\Psi\rangle$ is uniquely determined by the corresponding $\cal{P}$-context. This means that the probability
densities $|\Psi(x,t)|^2$ and $|\Psi(p,t)|^2$ to describe the $x$- and $p$-distributions of particles of this
ensemble depend, too, only on this $\cal{P}$-context; here $\Psi(x,t)$ and $\Psi(p,t)$ are the Fourier-transforms
of each other.

The $p$-distribution $|\Psi(p,t)|^2$ is evident cannot be reconstructed from the $x$-distribution
$|\Psi(x,t)|^2$, and vise versa. These distributions, associated with the same $\cal{P}$-context, supplement each
other, giving a complete information laid in the state-vector $|\Psi\rangle$ about the context and corresponding
quantum ensemble. The probability space to underlie the state-vector $|\Psi\rangle$ (or, the pair of statistical
distributions $|\Psi(x,t)|^2$ and $|\Psi(p,t)|^2$) is a non-Kolmogorovian one, irreducible in principle to the
sum of two Kolmogorovian subspaces \cite{Hr1} (it is relevant to note that these distributions are connected by
the uncertainty relation).

We have to stress once more that the statistical data described by $|\Psi(x,t)|^2$ and $|\Psi(p,t)|^2$ depend on
the $\cal{P}$-context, rather than on the $\cal{M}$-contexts under which the $x$ and $p$ observables are
measured. Quantum mechanics implies that the only role of any measurement is to reveal these two
($\cal{P}$-context's) probability distributions.

The generalized probabilistic approaches \cite{Hr5} and \cite{Ac5} imply that, for a single particle in a single
experiment, the values of the conjugate variables $x$- and $p$ can be measured under different
$\cal{M}$-contexts. As regards quantum mechanics, it deals with ensembles and says nothing about measurement. For
the ensemble created by some $\cal{P}$-context, Heisenberg's uncertainty relation does not forbid testing the
$x$- and $p$-distributions under the same $\cal{M}$-context (see \cite{Bal}). This relation reflects the
properties of $\cal{P}$-contexts, rather than $\cal{M}$-contexts. By Ballentine \cite{Bal}, for the principle
based on this relation, "[a] term such as {\it the statistical dispersion principle} would be really more
appropriate \ldots than the traditional name, {\it uncertainty principle}".

What indeed contradicts the uncertainty relation is the introduction of a joint $(x,p)$-distribution. As it
follows from this relation, there are correlations between these two conjugate variables (the origin of these
correlations can be revealed only within a prequantum theory), and, thus, they cannot be considered as mutually
independent ones.

Thus, in quantum mechanics, it is of secondary importance how the sets of statistical $x$- and $p$-data were
obtained, under different $\cal{M}$-contexts or not. Of more importance is that to introduce a joint
$(x,p)$-distribution for these sets is meaningless. In the case of ignoring this restriction, {\it different
observers} dealing with experimental $(x,p)$-data obtained for the same $\cal{P}$-context (i.e., for the same
ensemble) and, moreover, under the same $\cal{M}$-context will obtain {\it different $(x,p)$-"distributions"}.
But all they will recover from their $(x,p)$-data the same $x$- and $p$-distributions.

\subsection{The supplementarity principle for the wave and corpuscular properties of a particle} \label{a45}

The above approaches to criticize Bell's model of LHVs as well as to develop the concepts of contextuality and
supplementarity are essential for solving the quantum-classical problem. (We have also to stress the great
service of Bell who first used for studying this problem the inequality of probability theory.) Firstly, they rid
of obstacles the road for LHVs from the macro-world to the micro-world. Secondly, they discard Bohr's idea of
mutual incompatibility of the one-particle properties described by the noncommuting observables $x$ and $p$.

However, they leave untouched the contradiction to appear, in the standard description of CSMDSs, between the
quantum-mechanical superposition principle and probabilistic "either-or" rule. As before, the wave and
corpuscular properties of a particle look as mutually exclusive, and the conflict between quantum mechanics and
classical physics, exposed in the Schr\"odinger cat paradox, remains unresolved.

At this point it is important to stress that the adherence to the statistical (ensemble's) interpretation of
quantum mechanics (which is considered here as the only interpretation to reflect properly the nature of quantum
theory) does not free us from the necessity to solve this problem. This is so, because the contemporary
quantum-mechanical description of CSMDSs, which creates this problem, does not allow a consistent statistical
interpretation.

Indeed, as was shown in \cite{Ac1} by the example of a two-slit diffraction, the squared modulus of a CSMDS, in
any representation of this pure state, describes statistical data to belong to a non-Kolmogorovian probability
space. That is, if $|\Psi\rangle$ is a CSMDS to describe a one-particle process, then the probability space to
underlie, e.g., the corresponding $x$-distribution is non-Kolmogorovian (we stress that we do not speak here of
the non-Kolmogorovian probability space to underlie the whole state-vector $|\Psi\rangle$).

By probability theory (see \cite{Hr1}), a non-Kolmogorovian probability space describes {\it incompatible}
statistical data. For us this fact means that the squared modulus of a CSMDS cannot be unambiguously treated as
the probability density, and hence quantum mechanics itself, with its conventional approach to CSMDSs, does not
allow a correct statistical interpretation.

The main merit of our approach to CSMDSs is that it shows, by the example of two one-particle phenomena, that
quantum mechanics allows decomposing the non-Kolmogorovian probability space to underlie the squared modulus of a
CSMDS into the sum of Kolmogorovian ones. It suggests that, in the case of a CSMDS, the concept of
$\cal{P}$-contexts should be generalized. Namely, the $\cal{P}$-context to create a CSMDS should be considered as
a {\it complex} $\cal{P}$-context consisting of several {\it elementary} $\cal{P}$-contexts to create the
(sub)ensembles of quantum systems taking part in alternative subprocesses. For example, in the case of a 1D
completed scattering, we meet two elementary $\cal{P}$-contexts associated with transmission and reflection; in
the case of a two-slit diffraction either of two slits creates its own elementary $\cal{P}$-context and
alternative subprocess.

So, any quantum process described by a CSMDS must be considered as a complex one to consist of alternative
subprocesses associated with elementary $\cal{P}$-contexts. The squared modulus of a CSMDS, in any
representation, cannot be treated as the probability density, while the squared modula of its substates to
describe alternative subprocesses allow Born's interpretation. Of importance is to stress that elementary
$\cal{P}$-contexts are integral parts of the whole complex $\cal{P}$-context to be the "calling card"\ of the
phenomenon under study. This implies the inseparability of the corresponding subprocesses.

Note, our approach to CSMDSs implies also that, for any complex $\cal{P}$-context, there should exist a
counterpart which would, on the one hand, create an equivalent ensemble (described by the same state-vector
$|\Psi\rangle$), and, on the other hand, allow one to distinguish subprocesses associated with different
elementary $\cal{P}$-contexts. Such a counterpart will be referred to as the "which-way" $\cal{P}$-context.

For example, in the case of a 1D completed scattering, the which-way $\cal{P}$-context can be obtained using the
well-known Larmor procedure to imply the study of the spin's dynamics of an electron under an infinitesimal
magnetic field (see, e.g., \cite{But}). In the above model of a two-slit diffraction, the which-way
$\cal{P}$-context is associated with the modified experiment with a mirror. (We have to remark that, in the
two-slit experiment \cite{Afs}, the used which-way $\cal{P}$-context gives the interference pattern to differ
from the original one. Thus the reference to this modified two-slit experiment in the analysis of the original
one is irrelevant, as this step can lead to incorrect conclusions.)

So, our approach to CSMDSs reconciles the superposition principle with the above mentioned "either-or" rule. Now
the abbreviation 'CSMDS' can be equally applied to quantum and classical particles, and hence, in the
Schr\"odinger's cat paradox, the long-suffering cat is {\it either} died {\it or} alive, independently of an
observer, because at any instant of time the radioactive nucleus whose state evolves under a given complex
$\cal{P}$-context {\it either} has already decayed {\it or} has yet non-decayed.

\subsection{Some additional remarks on the concept of reality in a new approach to CSMDSs} \label{context}

As is widely accepted, within the statistical interpretation of quantum mechanics the state-vector corresponds to
nothing in the physical world. However, the generalized probabilistic approaches \cite{Hr5,Ac5} disprove this
viewpoint when the state-vector is not a CSMDS. Moreover, now we can say that, even if the state-vector
represents a CSMDS, this viewpoint is not true too. One has only to bear in mind that the squared modulus of a
CSMDS, in any representation, is not the probability density. The quantum process described by a CSMDS is a
complex one. In this case, only the squared modulus of each CSMDS's substate describes mutually compatible
statistical data, and hence reflects real (predetermined) properties of the subensemble moving under the
corresponding elementary $\cal{P}$-subcontext.

So, namely the squared modula of the CSMDS's substates -- "Kolmogorovian" statistical distributions -- describe
real physical properties of ensembles created by complex $\cal{P}$-contexts. In this case each observable should
be endowed with an additional index to specify the corresponding elementary $\cal{P}$-context (see also
\cite{Joy,Hes,Hr6} about the {\it realistic} Bell's type inequalities). For example, for a 1D completed
scattering the pair of the supplementary variables $x$ and $p$ splits into two pairs ($x_{tr}$, $p_{tr}$) and
($x_{ref}$, $p_{ref}$). For a two-slit diffraction we have two pairs ($x_{two}^{(1)}$, $p_{two}^{(1)}$) and
($x_{two}^{(2)}$, $p_{two}^{(2)}$).

\section{Conclusion}

We develop a new quantum-mechanical approach to a 1D completed scattering and two-slit diffraction. Each process
is presented as that consisting of two alternative subprocesses. The only exclusion appears in the case of a 1D
completed scattering, for wide (in the momentum space) wave packets, in the very stage of the scattering event.
We consider that this approach must be extended in quantum mechanics onto all quantum processes described by
CSMDSs. With the current approach to CSMDSs quantum theory does not allow a correct statistical interpretation,
because the squared modulus of a CSMDS describes mutually incompatible statistical data and, thus, cannot be
unambiguously interpreted as the probability density. At the same time the squared modula of the CSMDS's
substates to describe alternative subprocesses allow Born's interpretation.

We extend onto CSMDSs the concepts of contextuality and supplementarity proposed earlier within the generalized
probabilistic approach and present the program of solving the quantum-classical problem for isolated systems. The
above exclusion does not prevent the quantum-to-classical transition. The complete resolution of the
quantum-classical problem implies revising all the current models of quantum phenomena where CSMDSs appear.

\section{Acknowledgments}

The author thanks V.G. Bagrov and E. Recami for useful discussions of some questions of the paper. This work was
supported (in part) by Russian Science and Innovations Federal Agency under contract No 02.740.11.0238 as well as
by the Programm of supporting the leading scientific schools of RF (grant No 3558.2010.2).

\end{document}